\documentclass[iop,apj]{emulateapj}
\usepackage{apjfonts} % Times fonts

\usepackage{amsmath,amssymb,isomath,gensymb,graphicx,twoopt,lineno,color,soul}
\usepackage{hyperref}
\usepackage[maxfloats=50]{morefloats}

\sethlcolor{yellow}

\received{August 15, 2018}
\accepted{December 3, 2018}
\submitted{}
\journalinfo{Astrophysical~Journal}
\def\colhead#1{\multicolumn{1}{c}{\vrule depth 6pt width
0pt\relax#1}\ignorespaces}
\def\twocolhead#1{\multicolumn{2}{c}{\hss\vrule depth 6pt width
0pt\relax#1\hss}\ignorespaces}

\newcommand{\apla}[1]{{#1}}
\newcommand{\aplc}[1]{{#1}}
\newcommand{\yykb}[1]{{#1}}

\newcommand{\Gaia}{\textit{Gaia}\ }
\renewcommand{\vec}{\mathbfit}
\graphicspath{{./fig/}}

\shorttitle{Dissecting the AGN with VLBI \& Gaia}
\shortauthors{Plavin, Kovalev, \& Petrov}

\begin{document}

\title{Dissecting the AGN disk-jet system with joint VLBI-Gaia analysis}

% \correspondingauthor{A.~V. Plavin}
% \email{alexander@plav.in}

\author{
A.~V. Plavin\altaffilmark{1,2},
Y.~Y. Kovalev\altaffilmark{1,2,3},
\and
L.~Y. Petrov\altaffilmark{4,2}
}
\altaffiltext{1}{Astro Space Center of Lebedev Physical Institute, Profsoyuznaya 84/32, 117997 Moscow, Russia; alexander@plav.in}
\altaffiltext{2}{Moscow Institute of Physics and Technology, Dolgoprudny, Institutsky per. 9, Moscow region, 141700, Russia}
\altaffiltext{3}{Max-Planck-Institut f\"ur Radioastronomie, Auf dem H\"ugel 69, 53121 Bonn, Germany}
\altaffiltext{4}{Astrogeo Center, 7312 Sportsman Dr., Falls Church, VA 22043, USA}

\begin{abstract}
We analyze differences in positions of active galactic nuclei between \Gaia data release~2 and VLBI and compare the significant VLBI-to-\Gaia offsets in more than 1000 objects with their jet directions.
\aplc{Remarkably} at least 3/4 of the significant offsets are confirmed to occur downstream or upstream the jet representing a genuine astrophysical effect.
\aplc{Introducing redshift and \Gaia color into analysis can help distinguish} between the \aplc{contribution} of the host galaxy, jet, and accretion disk emission.
\yykb{We find that strong optical jet emission at least 20-50pc long is required to explain the \Gaia positions located downstream from VLBI ones}.
Offsets \aplc{in the upstream direction} of up to 2~mas are at least partly due to the dominant impact of the accretion disk on the \Gaia coordinates and by the effects of parsec-scale radio jet.
The host galaxy was found not to play an important role in the detected offsets.
BL~Lacertae object and Seyfert~2 galaxies are observationally confirmed to have a relatively weak disk and consequently downstream offsets.
The disk \aplc{emission drives upstream offsets in a significant fraction of quasars and Seyfert~1 galaxies when it dominates over the jet in the optical band}.
The observed behaviour of the different AGN classes is consistent with the unified scheme assuming varying contribution of the obscuring dusty torus and jet beaming.
\end{abstract}


\section{Introduction}

Both VLBI and Gaia \citep{r:Gaia} provide positions for various objects, including active galactic nuclei (AGNs) with sub-milliarcsecond accuracy. Comparison of VLBI positions with those from \Gaia data release~1 (GDR1) revealed that 7\,\% of AGNs have statistically significant differences that cannot be explained by random measurement errors \citep{r:gaia1}.
Subsequently, we discovered that the directions of VLBI-to-\Gaia position offsets are not isotropic and have a strong concentration downstream and upstream the jet \citep{r:gaia2}. The discovery presented a strong evidence that these position differences are not artifacts of radio or optical data analysis, but are due to the structure of the sources.
We suggested in \cite{r:gaia2,r:gaia3} that the presence of strong parsec-scale optical jet \aplc{emission} causes these offsets.
We predicted that further improvements in the VLBI and \Gaia accuracy will not reconcile position differences, and even more sources will have accurately measured significant offsets.
We note that astrophysical properties of active galactic nuclei at parsec scales are still sometimes not assumed as playing the major role in significant VLBI-\Gaia offsets \cite[e.g.][]{r:Gaia-CRF2}. \cite{2018AJ....155..229F} discussed significant VLBI-\Gaia offsets for 20 sources and attributes the majority of them to binary or extended objects.

The \Gaia data release~2 \citep[GDR2,][]{r:GDR2cat} brought a substantial improvement with respect to the GDR1: the number of sources increased by almost 50\,\%, the position accuracy of VLBI-\Gaia matched AGNs improved by a factor of 5, and magnitudes at red and blue passbands became available. This prompted us to update and expand the previous analysis, check our predictions, and study the origin of these offsets in more detail using additional information.

\begin{figure*}
    \centering
    \includegraphics[width=0.95\textwidth,trim=0cm 0cm 0cm -0.3cm]{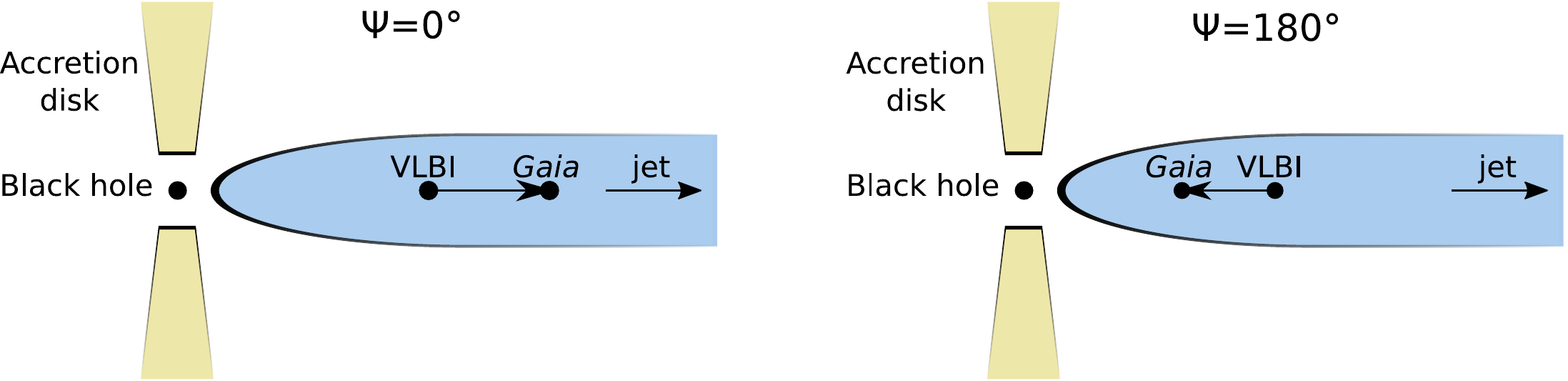}
    \caption{A cartoon explaining two opposite VLBI-to-\Gaia offset directions \aplc{with respect to the parsec-scale jet: downstream $\Psi = 0\degr$ and upstream $\Psi = 180\degr$}.}
    \label{f:diagram}
\end{figure*}

\begin{figure*}
    \centering
    \includegraphics[width=\textwidth,trim=0cm 0.3cm 0cm -0.2cm]{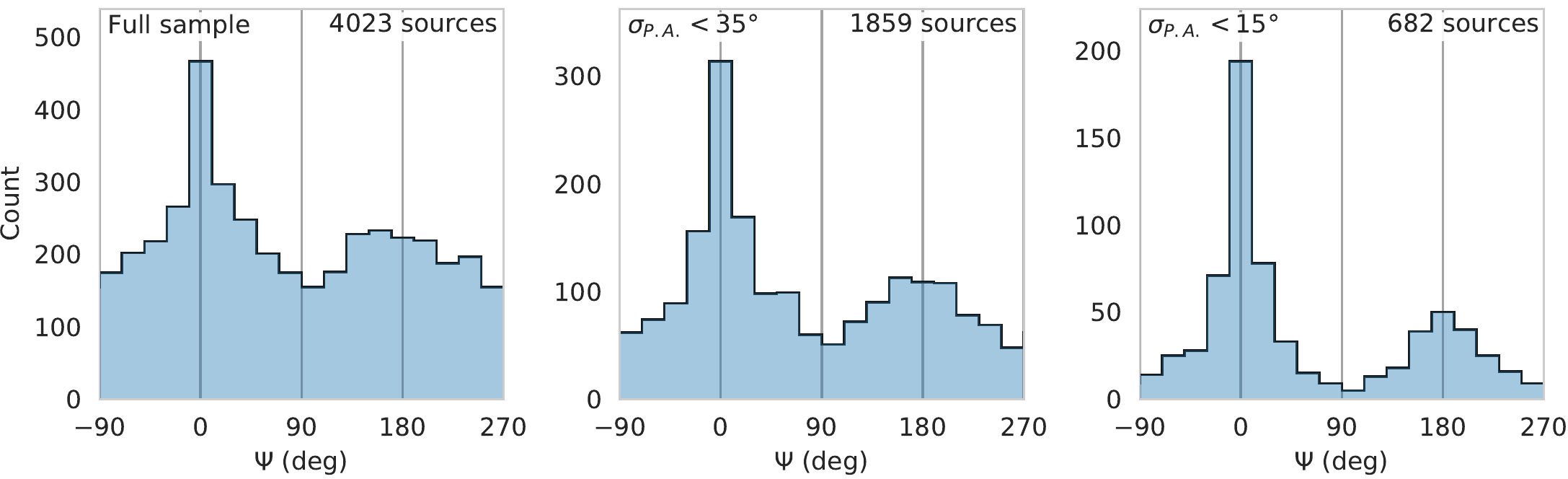}
    \caption{Distribution of the offset direction relative to the jet, $\Psi$, for the whole set of sources and for two subsets filtered by restricting to small angle error estimates.}
    \label{f:ltr_hists}
\end{figure*}

Throughout this paper we define and use the offset $\vec{VG}$ as the vector from VLBI to \Gaia coordinates of a source. Its direction is described by the angle $\Psi$ relative to the parsec-scale jet direction, see the cartoon in \autoref{f:diagram}.
The \Gaia position is shifted downstream the jet relative to VLBI when $\Psi\approx0\degree$, i.e.\ the optical centroid is farther away from the nucleus.
The \Gaia position is shifted upstream the jet relative to VLBI position when $\Psi\approx180\degree$, i.e.\ the optical centroid is closer to the nucleus than the VLBI position.

We consider the optical emission of an AGN as coming from three main components: the jet, the accretion disk, and the host galaxy.
\Gaia coordinates derived from CCD measurements correspond to the optical centroid of these components.
VLBI coordinates derived from correlated visibility data correspond not to the radio-band centroid position, but to the most compact feature.
\apla{The difference in the measurement principles implies that even if radio- and optical structure of AGNs are similar and colocated, the coordinates measured by VLBI and \Gaia will still be different  \citep{r:gaia3}.}
The most compact feature typically coincides with the so-called radio core which is offset by less than 1~mas downstream the jet from its origin due to synchrotron opacity \citep{Kovalev_cs2008,r:Porcas09,r:MOJAVE_cs,r:cs_var}.
\apla{Different aspects} of this three-component interpretation are further tested using the difference of \aplc{optical} spectra of the jet, accretion disk, and host galaxy.

\section{Observational data}
\label{s:data}

We cross-identify the VLBI-based Radio Fundamental Catalogue\footnote{\url{http://astrogeo.org/vlbi/solutions/rfc_2018b}} (RFC) with the \Gaia Data Release~2 as we did for GDR1 in \cite{r:gaia1}. For the following analysis, we select 9081 matched sources with \apla{per-source} probability of false association $\mathrm{PFA}<2\cdot10^{-4}$. The median VLBI-\Gaia offset is 1.4~mas and the maximal is 400~mas \citep[see for details][]{r:gaia4}.
We supplement the \Gaia positional information by the \apla{optical color \aplc{indices} derived from three-band \Gaia photometry \citep{r:GDR2passband} with approximate ranges 330-1000 nm for G, 330-660 nm for BP, and 630-1000 nm for RP}.
While \cite{r:gaia4} showed that VLBI and \Gaia position errors are understimated, the values reported in the catalogues are used in our analysis without any scaling. Taking this scaling into account does not change any of our conclusions.

The parsec-scale jet directions are estimated using VLBI images collected in the Astrogeo database\footnote{\url{http://astrogeo.org/vlbi_images/}}. The images come from various surveying programs as detailed in \cite{r:gaia2}. Images with frequency bands ranging from 1.4 to 40~GHz are used in the analysis. For sources having more than one image available we use a median jet position angle. We succeeded in determining the jet direction in 4023 AGNs among the RFC-GDR2 matches.
In addition to RFC and GDR2 catalogues, we take the redshifts and classes of AGNs from the NASA Extragalactic Database (NED).

\section{Dominant reason of the significant VLBI-\Gaia positional offsets}
\label{s:reason}

\begin{figure*}
    \centering
    \includegraphics[width=\textwidth,trim=0cm 0cm 0cm 0cm]{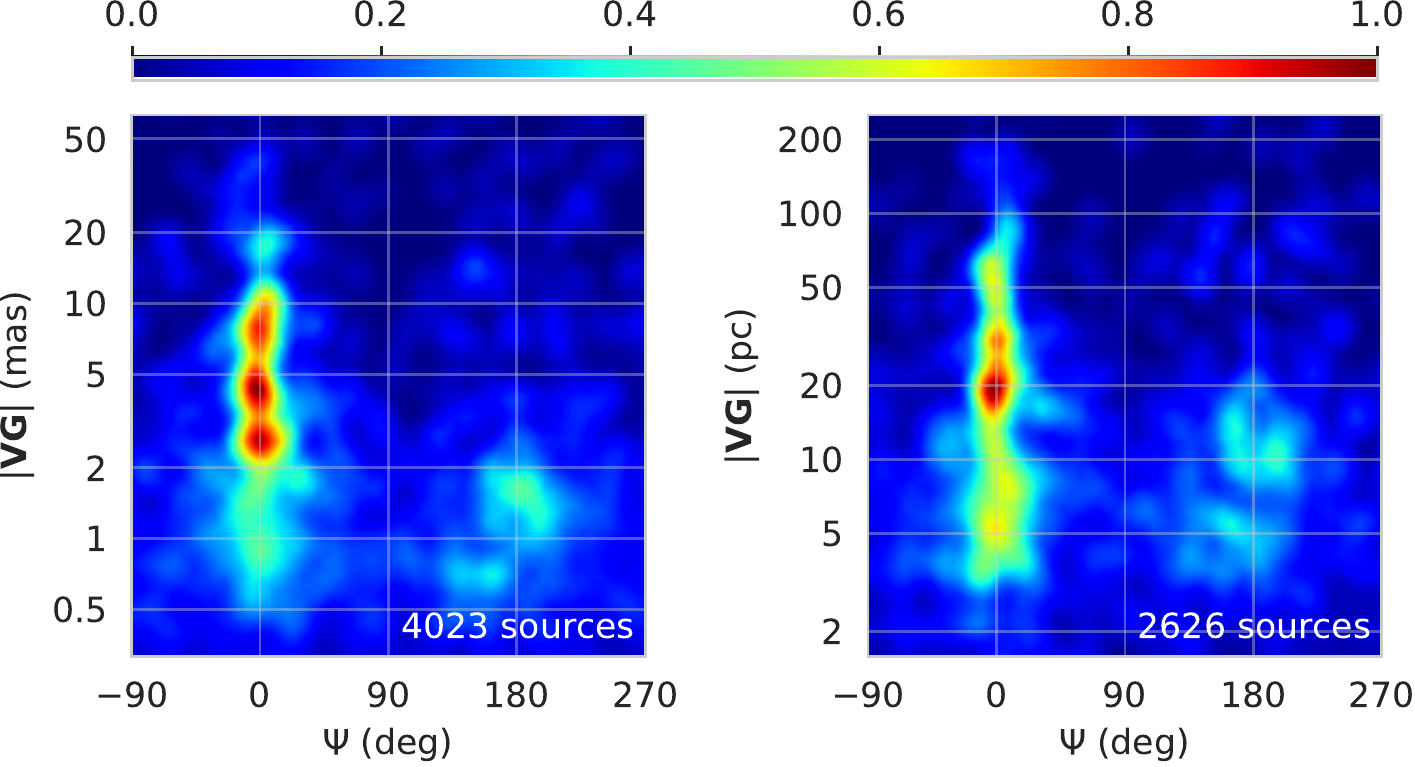}
    \caption{Pairwise distribution of VLBI-\Gaia offset direction and length in projected angular (left) and linear (right) units. \apla{Here and below we show a weighted linear kernel density estimate \citep{doi:10.1080/03610929808832217}}.
    \label{f:ltr_pa_vg}}
\end{figure*}

The distribution of the VLBI-\Gaia offset-jet angles $\Psi$ is given in \autoref{f:ltr_hists}. Histograms are shown separately for the whole set of matching sources and for two subsets filtered by the uncertainty of $\Psi$. We compute it assuming that the distribution of measurement errors for both VLBI and \Gaia are zero-mean Gaussians with standard deviations and correlations reported in the corresponding catalogues. The filtering thresholds we use, namely $\sigma_\Psi$ less than $35\degree$ and $15\degree$, roughly correspond to a $2\sigma$ and $4\sigma$ cutoff for a 2D Gaussian.

The anisotropy of $\Psi$ can clearly be seen on all three of the histograms in \autoref{f:ltr_hists} and it is more pronounced than using GDR1 \citep{r:gaia2}. This confirms our original conclusions: the VLBI-\Gaia offsets have two preferred directions, for $\Psi\approx0\degree$ downstream and $\Psi\approx180\degree$ upstream the parsec-scale radio jet. The effect is real, it was not caused by some unaccounted effects in the GDR1 or GDR2 data.
The peaks at 0\degr\ and 180\degr\ in the $\Psi$ distribution can happen only if positions are predominantly affected by the jet. Thus, the $\vec{VG}$ position angles for sources where the significant offset is unrelated to the jet or caused by unaccounted measurement errors are distributed uniformly.
We exploit this property for counting the fraction of objects with VLBI-to-\Gaia position offsets caused by the jet. We count sources with $\Psi$ in the range $90\degr\pm20\degr$ and $270\degr\pm20\degr$ and take this as an upper bound of the density of sources with an offset not dominated by the jet. Then, subtracting the count $N_1$ corresponding to that density from the total number of sources $N$ we get a lower bound on the number of jet-affected AGNs: $N-N_1$; the lower bound on the fraction of such sources is given by $1-N_1/N$. We \aplc{find $950\pm150$ AGNs with} offsets significantly affected by the jet in the whole dataset (left in \autoref{f:ltr_hists}). However, many low-significance offsets are dominated by the position uncertainties and contribute to our estimation by smearing the $\Psi$ distribution. So, we repeat this analysis for sources with $\sigma_\Psi<15\degree$ (right in \autoref{f:ltr_hists}) and get that $475\pm40$ or a fraction of ($73\pm6$)\,\% of them have offsets dominated by the jet.
We conclude that \aplc{at least 73\,\% of sources with significant VLBI-\Gaia position differences have offsets driven by relativistic jets, leaving less than 27\,\% with randomly oriented offset directions}.
\cite{r:gaia4} have modeled this distribution differently, under stronger model assumptions, but achieved similar results.
\apla{We propose extended optical jets as the explanation of the 0\degr-offsets downstream the jet. \aplc{See also discussion in \citet{r:gaia2} who analysed \Gaia DR1 data.} Host galaxy and dusty torus effects are discussed in \autoref{s:host}.}

As the GDR2 contains more sources and their coordinates are more accurate compared to GDR1, we can analyze the joint distribution of offset-jet angle $\Psi$ and the offset length $|\vec{VG}|$ (\autoref{f:ltr_pa_vg}).
Note that instead of filtering sources using a $\sigma_\Psi$ threshold, we apply an error-based weighting while including all the sources. Specifically, each source has a weight $w = 1/\sqrt{\sigma_\Psi^2 + (5\degree)^2}$ where the $5\degree$ term is chosen empirically and accounts for jet direction uncertainty as well as other sources of error. 
%\apla{We use the same weighting and a linear kernel for all bivariate plots in this paper}.

As one can see, sources with the \Gaia position further down the jet from the VLBI one, are present for basically any offset length up to 50~mas or $\sim200$~pc projected distance. \aplc{There are 365 sources with significant offsets in this direction longer than 1~mas, and they constitute $9.1$\,\% of our sample.}
\aplc{These position differences are consistent with} our interpretation that 0\degr-offsets are primarily caused by bright and extended optical jets \citep{r:gaia2,r:gaia3}. Lengths of offsets in the $\Psi=0\degr$ directions imply that 20-50~pc optical jets are quite common while some of them extend even beyond 100~pc.

Sources with the 180\degr-offsets are concentrated at smaller $|\vec{VG}|$, less than 2~mas or 20~pc. 
This requires that the VLBI position is shifted downstream from the central engine.
\aplc{The unaccounted source structure contribution to group delay and 
frequency-dependent synchrotron opacity (core-shift) may cause a shift in the 
estimates of radio positions towards that direction. However, the typical 
magnitude of this shift is estimated to be at a level of 0.2~mas 
\citep{Kovalev_cs2008,r:Porcas09,r:gaia3}, i.e. about one order of magnitude 
smaller than observed by us. A large fraction of AGNs is expected to have 
upstream VLBI-\Gaia offsets which are not seen at the current level of 
positional precision. We expect that they will appear in the next VLBI and \Gaia 
data releases. Note that 138 sources with significant upstream VLBI-\Gaia DR2 
offsets longer than 1~mas constitute only $3.4$\,\% of our sample, so they 
certainly do not represent the typical case. We surmise that offset values about 
1.5~mas might represent the tail of their distribution partly affected by the 
core shift variability \citep{r:cs_var} and/or the magnitude of the contribution 
of source structure and core-shift is significantly underestimated and/or there 
is another, yet unknown, cause of the offsets at the $180^\circ$ direction. We plan to investigate these hypotheses in detail in the future.}
Additionally, the \Gaia centroid of these objects should point close to the central engine position either due to the dominance of the accretion disk or the optical jet base. These scenarios will be examined in the next section in detail. Since the radio sky is dominated by one-sided jets due to \aplc{Doppler} boosting, we do not consider counter-jets.

\begin{figure*}
    \centering
    \includegraphics[width=\textwidth,trim=0cm 0cm 0cm 0cm]{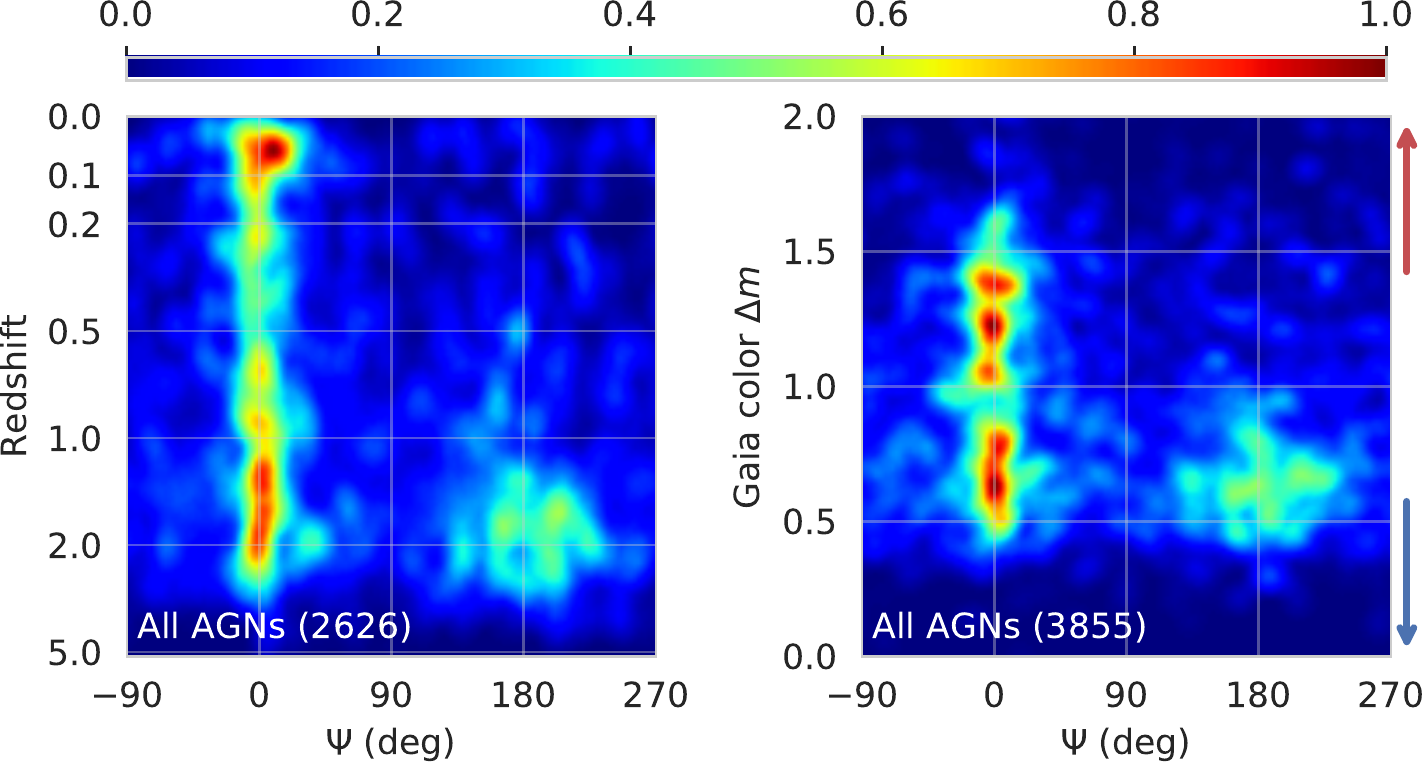}
    \caption{Pairwise distribution of $\Psi$ together with redshift (left) and optical color \aplc{index $\Delta m$} (right). The color \aplc{index} is defined as the difference of \Gaia magnitudes $\Delta m = m_\text{BP} - m_\text{RP}$, so the \yykb{top of this plot corresponds to the redder sources while the bottom to the bluer ones}.}
    \label{f:ltr_pa_rs_color}
\end{figure*}

\aplc{Since almost all images were generated using VLBA observations, their resolution along the declination axis is usually poorer than along the right ascension axis. VLBI coordinates also tend to have higher uncertainty along the declination axis. To ensure that the offset-jet alignment is not due to this disparity in resolution, we repeated our analysis by dropping the sources with jet position angle withing $\pm20^\circ$ of the declination and right ascension axes separately. Plots and results of the analysis do not differ qualitatively from the full sample presented in the paper. The RFC catalogue was formed to be complete down to at least 200~mJy at 8 GHz \citep{VCS5}. Repeating our analysis for this flux-limited subsample also confirms conclusions presented here.
No instrumental errors of \Gaia could lead to the preferred direction of the offsets being parallel to the jet. If the coordinates measured by the CCD matrix depend on the object color, it would lead to sources with uniformly distributed offset-jet angle, as the CCD bias can not coincide with the jet directions. We conclude that the results are robust to possible systematic effects of our instruments.
}

\section{AGN redshift and optical color as a tool to probe the disk-jet system}
\label{s:redshift_color}

\aplc{In this section we develop our explanation of $0\degree$ offsets as resulting from strong emission of optical jets and $180\degree$ offsets as being associated with accretion disk emission. We show how this explanation is supported by observational evidence, using optical color as a critical probe.}
Parsec-scale jets in low-synchrotron-peaked AGNs have falling synchrotron spectrum in the optical band while the accretion disks spectra peak in UV \citep{r:AGNreview2017}. Consequently, AGNs with dominant jets are expected to look redder in \Gaia band in comparison \aplc{to those where accretion disk dominates.} \aplc{More distant AGNs should exhibit higher apparent dominance of accretion disk for \Gaia due to the cosmology-driven frequency shift of UV emission into the \Gaia passband \citep{r:GDR2passband}. Correspondingly, one can expect that the \Gaia position of bluer or more distant objects is affected by the accretion disk more. Note that a source might be bluer either because the UV emission is redshifted into the \Gaia band, or because it has intrinsically more optical emission from the accretion disk.}

We show the distribution of the VLBI-\Gaia offset-jet angle $\Psi$ and the redshift in the left panel of \autoref{f:ltr_pa_rs_color}. It is clearly visible that the majority of 180\degr-offsets are observed at large redshifts $z \geq 1.5$ while the 0\degr-offsets are present for all redshifts starting from zero.
The right panel of \autoref{f:ltr_pa_rs_color} shows the distribution of the offset-jet angle and the color \aplc{index}, which we define as the difference of \Gaia magnitudes in `blue' ($m_\text{BP}$) and `red' ($m_\text{RP}$) bands. The majority of sources with the 180\degr-offset are relatively blue (bottom right corner of the plot), while the 0\degr-offsets occur for sources of different colors.
\aplc{This suggests} that VLBI-\Gaia offsets \aplc{in the upstream direction} predominantly happen when \aplc{the bluer accretion disk emission is more prominent in the \textit{Gaia} band.}
\yykb{We note that the redder objects show 0\degr-offsets only. This strongly supports the optical jet hypothesis for the downstream offsets.}

\section{AGN classes with different host galaxy -- accretion disk -- jet contributions}
\label{s:classes}

\begin{deluxetable*}{cccccc}
\tablecaption{General information about VLBI-\Gaia matched AGNs.}
\tablehead{\twocolhead{Optical class\tablenotemark{a}} & \colhead{Number} & \colhead{Median redshift\tablenotemark{a}} & \twocolhead{Median VLBI-\Gaia offset\tablenotemark{b}} \\ 
\colhead{} & \colhead{} & \colhead{} & \colhead{} & \colhead{(mas)} & \colhead{(pc)} } 
\startdata
Quasars  &  &          1891 &  1.40 &  0.70 &  5.5 \\
BL Lacs  &  &           347 &  0.42 &  0.62 &  3.4 \\
Seyferts & type 1 &     127 &  0.34 &  0.73 &  3.1 \\
         & type 2 &      25 &  0.15 &  7.2 &   15  \\
         & unknown &     37 &  0.26 &  1.5 &   4.6 \\
Other\tablenotemark{c}    &  &           437 &  0.21 &  2.3 &   7.1 \\
Unknown  &  &          1159 &   &  1.4 &           \\
\enddata
\tablenotetext{a}{Optical classes and redshifts are based on the NED. Note that redshift information is available for about 90\,\% of the AGNs with known class.}
\tablenotetext{b}{Offsets are given in projection on the sky plane.}
\tablenotetext{c}{The ``Other'' subset is dominated by radio galaxies.}
\label{tbl:general}
\end{deluxetable*}

\begin{figure}
    \centering
    \includegraphics[width=1.0\columnwidth,trim=0cm 0cm 0cm 0cm]{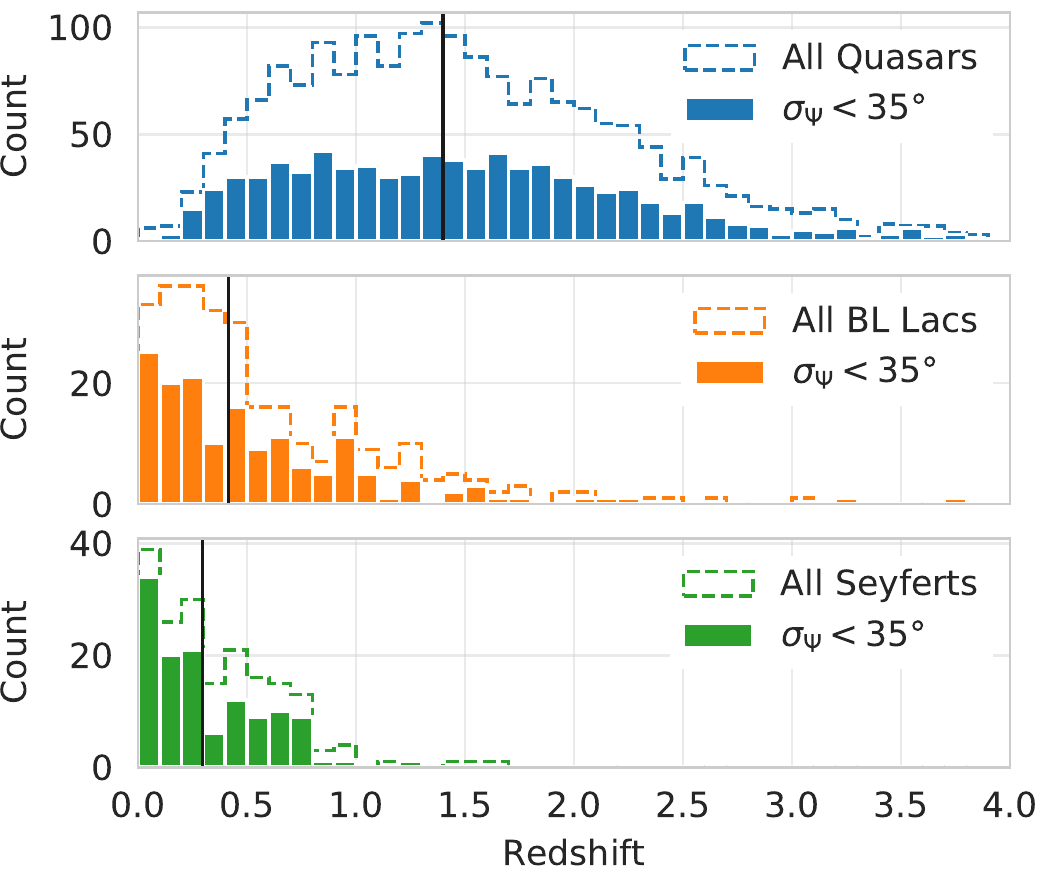}
    \caption{Distribution of redshift for different AGN classes. The median values are shown by vertical lines and are the following: $z=1.4$ for quasars, $z=0.4$ for BL Lacs, $z=0.3$ for Seyferts.}
    \label{f:ltr_redshift_hist}
\end{figure}

\begin{figure*}
    \centering
    \includegraphics[width=0.96\textwidth,trim=0cm 0cm 0cm 0cm]{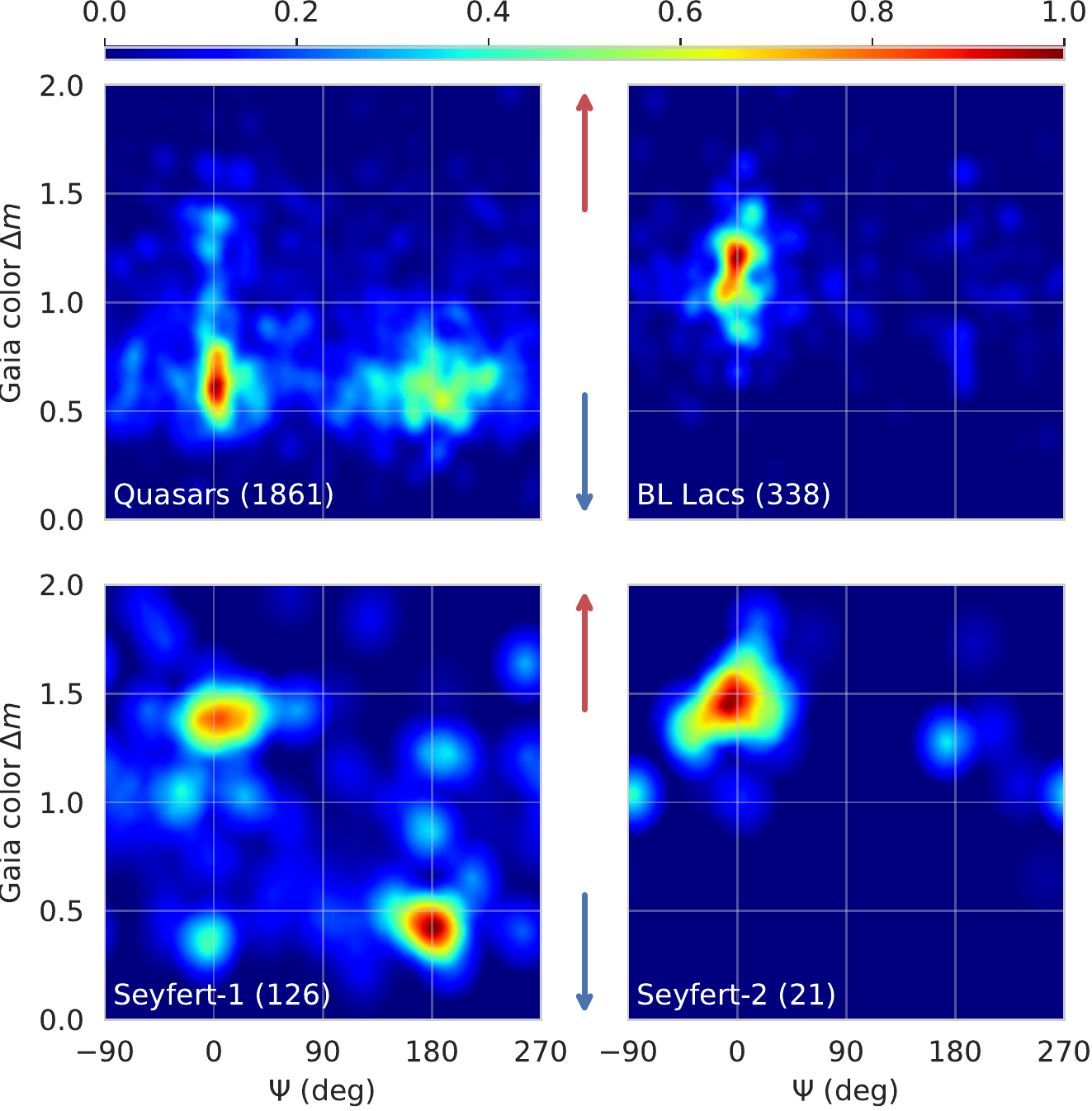}
    \caption{Pairwise distribution of $\Psi$ together with optical color \aplc{index $\Delta m$} shown separately for different classes of AGNs.}
    \label{f:ltr_pa_color_type}
\end{figure*}

Various classes of AGNs have significantly different contributions of their relativistic jets, accretion disks, and host galaxies to the observed emission. Studying the VLBI-\Gaia offsets for each class separately can provide an additional test for our interpretation of the offsets and lead to new insights about inter-class differences. This includes testing predictions of the unification scheme which implies that observing an AGN from different angles determines the level of obscuration of the nucleus by the dusty torus and emission of the beamed jet \citep{r:Antonucci93,r:UP95,r:vcv2000,1989ApJ...336..606B}. \aplc{We focus our study on quasars, BL~Lac objects and Seyfert galaxies (type~1 and type~2).
}
For definition of the AGN classes \apla{see \cite{VCV2010}}.
Typical redshifts of objects differ between classes, as shown in \autoref{f:ltr_redshift_hist}: in the increasing order of distance they are Seyferts, BL~Lacs, and quasars. \aplc{Summary of sources for each AGN class is given in Table~\ref{tbl:general} with VLBI-\Gaia offset values not filtered by significance.}

We show the joint distribution of the offset-jet angle $\Psi$ and color \aplc{index} for each of these classes in \autoref{f:ltr_pa_color_type} in the same way as in the right panel of \autoref{f:ltr_pa_rs_color}. One can clearly see significant differences between the classes. Source fraction estimates given below include only objects with $\sigma_\Psi < 35\degree$. The general trend that the \aplc{upstream} offsets are present mostly for bluer objects remains even when we consider classes separately.
We analyzed sub-samples of quasars, BL Lacs, Seyfert galaxies, and radio galaxies with similar redshifts. The conclusions presented in this section have been confirmed. This means that the discussed differences between the classes are not due to different redshifts only.

The majority of \textit{quasars} have color \aplc{index} $\Delta m \approx 0.6$, and only 20\,\% of them are redder than $\Delta m > 1$. More than 2/3 of these red ones have downstream 0\degr-offsets, and only 14\,\% of quasars with  upstream 180\degr-offsets are red. Quasars with $\Delta m \approx 0.6$, which have offsets both downstream (48\,\%) and upstream (52\,\%), most likely have significant contribution of both accretion disk and jet in the optical band. In this case the offset length and direction depend on the disk-to-jet flux ratio and on the jet length. Moreover, the offset direction for a given object may flip in time depending on the activity state of the disk and jet. Detailed analysis of corresponding effects will be possible with further releases providing data on temporal \Gaia positions and photometry changes.

\textit{BL~Lac objects} have no or very faint broad line emission. This can be explained if the accretion disk is weak and can not excite the broad line region \citep{G11} or the contribution of the highly beamed continuum emission of the jet covers the lines \citep{r:UP95}.
If the disk is what causes the 180\degr-offset upstream the jet, only a few BL Lacs should exhibit such offsets. This is exactly what we see in \autoref{f:ltr_pa_color_type}: 85\,\% of them have 0\degr-offsets downstream.
\aplc{This leaves optical emission of the accretion disk and not the jet base as the major cause of the 180\degr-offsets. Indeed, the nucleus and jet origin of BL~Lacs are not blocked by the dusty torus. These objects are observed face~on, and the bright base of the optical jet \citep[see modeling by][]{r:AGNjet_optical_spectrum} is visible.
Still, its emission turns out to be not strong enough to cause the 180\degr-offset when the contribution of the accretion disk is low.}

\textit{Seyfert galaxies} are relatively faint active galaxies with weak jets. A large fraction of optical emission comes from their host galaxies \citep{r:weedman1977}.
Seyfert galaxies are divided into two sub-classes significantly different for our purposes: type~1 having both narrow and broad emission lines, and type~2 with narrow lines only. Depending on the orientation of the object \citep{r:AM85}, the central \apla{region} is visible and dominant (type~1) or hidden behind the dusty torus (type~2). Therefore one can expect that Seyfert~1 \aplc{may} have VLBI-\Gaia offsets \aplc{upstream} the jet, but Seyfert~2 have almost none. Indeed, we found that more than 80\,\% of Seyfert~2 are offset downstream, while about a half (42\,\%) of Seyfert~1 have upstream offsets (\autoref{f:ltr_pa_color_type}).
We detect the strong relationship between the color of Seyfert galaxies and the offset direction: two thirds of objects fall into top-left or bottom-right corners in the diagrams. They are jet or accretion disk dominated, respectively. \aplc{Even though the nuclear extinction in Seyfert~1 objects is weaker compared to Seyfert~2, it still may partially obscure the central part depending on their orientation. Seyfert galaxies may look jet-dominated and have upstream offsets due to this extinction.}
The found $\Delta m$ values for Seyfert galaxies with offsets upstream and downstream the jet and their difference of about 1 constrain the optical spectra of the dominant nucleus (the disk plus jet base) and the rest of the optical jet. This result can also be useful to constrain dimensions of the dusty torus applying the measured offset lengths.

\apla{
\section{Host galaxy and obscuring torus}
\label{s:host}

The analysis of the distributions of $\Psi$ and colors allows us to constrain the host galaxy contribution to offsets. If the central parts of the hosts were asymmetric and unrelated to the jets, their centroids would shift the \aplc{VLBI-\Gaia} offsets in random directions. Our estimates in \autoref{s:reason} imply that \aplc{ no more than 27\,\% of significant offsets are not aligned with the jets}. Future \Gaia releases will constrain this upper limit further. \aplc{Note that this holds true even for objects with bright host galaxies, such as Seyferts.} If the galaxies were symmetric around the optical nucleus, they would cause offsets with $\Psi\approx 180\degree$. However, most of host galaxies have falling spectra \citep[see e.g.][]{2003ApJ...583..159H} corresponding to the red color, \aplc{while we see such offsets predominantly in bluer sources in \autoref{f:ltr_pa_rs_color} and \autoref{f:ltr_pa_color_type}.}

Extinction of host galaxy emission by the dusty torus may lead to jet-aligned optical structure of the objects \citep{2018MNRAS.475.5179S}. However, the majority of sources in our sample are blazars which have small viewing angles around 5\degr\ \citep{2010A&A...512A..24S}. For such angles the obscuration of the jet- and counter-jet sides of the host is basically the same, although the difference is hard to quantify without further assumptions.

Seyfert galaxies are objects with a large variety of viewing angles. \autoref{f:ltr_pa_color_type} presents two distinct groups: bluer with upstream and redder with downstream offsets. 
The very small accretion disk can be either completely blocked by the torus or completely visible leading to downstream and upstream offsets, correspondingly. This naturally explains the observed dichotomy.
If the offsets were due to shadowing of host galaxy emission by the torus, a continuous range of changing offset and color values would be observed due to the continuously changing level of obscuration.
}

\aplc{
\section{Summary}
\label{s:summary}

We confirm using \Gaia Data Release 2 that the \Gaia offsets with respect to VLBI position have predominant directions downstream or upstream the jet. This means the results of \cite{r:gaia2} represent a genuine astrophysical effect and were not caused by some unaccounted errors in DR1. We estimate that for at least $73$\,\% of sources with large offsets which have $\sigma_\mathrm{P.A.} < 15\degr$ the VLBI-\Gaia offsets are dominated by the effect of jets. This is a significant increase from our previous bound of 53\,\% based on \Gaia DR1, and we expect this fraction to increase further as the accuracy of VLBI and \Gaia measurements is continously improving. This means that the fraction of offsets caused by other effects in AGNs may be much lower than our current bound of 27\,\%. We find that the host galaxy emission which may partially be obscured by the dusty torus does not have a dominant effect on the VLBI-\Gaia offsets.

We explain VLBI-\Gaia offsets in the downstream direction by the presence of bright and extended optical jets which shift the \Gaia centroid. From the magnitudes of such offsets we infer that optical jets of projected length at least 20-50~pc are quite common among AGNs. 

VLBI-\Gaia offsets in the upstream direction indicate that the VLBI positions do not coincide with the nuclei confirming predictions by e.g.~\cite{Kovalev_cs2008,r:gaia3}. The VLBI positions are shifted by up to 2~mas or 20~pc projected downstream the jet.
Unaccounted structure of radio jets and frequency-dependent synchrotron opacity may contribute to these offsets. However, their contribution is expected typically at a level of 0.2~mas or less. We surmise that either the observed offsets represent the tail of the distribution, or the contribution of source structure and core shift is underestimated, or another, yet unknown effect, causes such large shifts.
Note that the typical expected shifts are shorter than 1~mas and can not yet be detected within the errors of VLBI and \Gaia positions.

For objects with upstream VLBI-\Gaia offsets the optical position should be very close to the nucleus. 
Such AGNs are found to be typically located at higher redshifts and are bluer than those with offsets downstream the jet. As the accretion disk mostly emits at UV wavelengths, both higher redshifts and bluer colors indicate that more of the accretion disk flux falls into the optical passband of \textit{Gaia} affecting the measured position.
We conclude that in the cases of upstream offsets \Gaia coordinates are mostly determined by the dominant emission of the accretion disk.

Examining quasars, BL Lacs and Seyfert galaxies separately we find more support for our claims and get new insights on the classification itself.
The unified scheme of active galaxies explains different AGN classes by orientation effects which involve Doppler boosting and obscuration by the dusty torus. It can also satisfactorily explain the found color index -- offset direction diagrams for different AGN classes.
Quasars and Seyfert~1 galaxies may shift in both directions depending on relative contributions of their disks and jets.
BL Lacs and Seyfert~2 galaxies have relatively weak BLR and accretion disk emission or the central region is obscured, and we find  that almost none of them have upstream VLBI-\Gaia offsets. 
This is a strong evidence that \Gaia coordinates correspond to position of the nucleus due to the strong accretion disk and not just the jet origin being brighter in the optical band.
}

\vspace*{0pt}
\acknowledgments
We deeply thank the teams referred to in \S~\ref{s:data} for making their fully calibrated VLBI FITS data publicly available as well as Richard Porcas, Eduardo Ros, Galina Lipunova and the referee Ian Browne for useful comments.
This research has made use of data from the MOJAVE database that is maintained by the MOJAVE team \citep{MOJAVE_XV}.
This study makes use of 43~GHz VLBA data from the VLBA-BU Blazar Monitoring Program, funded by NASA through the Fermi Guest Investigator Program.
This research was supported by Russian Science Foundation (project 16-12-10481).
This work has made use of data from the European Space Agency (ESA) mission \textit{Gaia}\footnote{\url{https://www.cosmos.esa.int/gaia}}, processed by the \Gaia Data Processing and Analysis Consortium (DPAC\footnote{\url{https://www.cosmos.esa.int/web/gaia/dpac/consortium}}). Funding for the DPAC has been provided by national institutions, in particular the institutions participating in the \Gaia Multilateral Agreement.
This research has made use of NASA's Astrophysics Data System.
This research has made use of the NASA/IPAC Extragalactic Database (NED), which is operated by the Jet Propulsion Laboratory, California Institute of Technology, under contract with the National Aeronautics and Space Administration.

\vspace{2mm}
\noindent
Facilities: \textit{Gaia}, VLBA, EVN.

\bibliographystyle{apj}
\bibliography{VLBI_GDR2_colors}

\end{document}